\documentclass[12pt,english]{article}
\usepackage{babel}
    \title{Number Operator Algebras and Deformations of $\epsilon$-Poisson Algebras}
    \author{Fabien Besnard\footnote{Universit\'e Paris 7, UFR de math\'ematiques, case 7012, 2 place Jussieu, 75251 Paris Cedex 05. e-mail : besnard@math.jussieu.fr}}

\def\CC{{\rm\bf C}}

\def\ZZ{{\rm\bf Z}}
\def\NN{{\rm\bf N}}
\def\II{{\cal I}}
\def\JJ{{\cal J}}
\def\KK{{\cal K}}

\def\SS{{\cal S}}

\def\fer{{\cal C}}
\def\bos{{\cal A}}
\def\aix{{a^+_i}}
\def\ai{{a^{}_i}}
\def\ajx{{a^+_j}}
\def\aj{{a^{}_j}}
\def\akx{{a^+_k}}
\def\ak{{a^{}_k}}

\def\ajj{{a^{}_\JJ}}
\def\akk{{a^{}_\KK}}
\def\aiix{{a^+_\II}}
\def\ajjx{{a^+_\JJ}}
\def\dem{{\bf Proof : }}
\def\Hom{\mbox{Hom}}
\def\hattens{\mathop{\hat{\bigotimes\kern 0pt}}\limits}
\def\tens{\mathop{\bigotimes\kern 0pt}\limits}
\def\ep{\epsilon}
\newtheorem{theorem}{Theorem}
\newtheorem{definition}{Definition}

\def\be{\begin{equation}}
\def\ee{\end{equation}}
\def\bea{\begin{eqnarray}}
\def\eea{\end{eqnarray}}

\makeatletter
\renewcommand{\@biblabel}[1]{\quad #1.}
\makeatother

\begin{document}
\maketitle
\begin{abstract}
It is well known that the Lie-algebra structure on quantum algebras gives rise to a Poisson-algebra structure on classical algebras as the Planck constant goes to 0. We show that this correspondance still holds in the generalization of super-algebra introduced by Scheunert, called $\ep$-algebra. We illustrate this with the example of Number Operator Algebras, a new kind of object that we have defined and classified under some assumptions.
\end{abstract}
\eject

\section{Introduction}
In \cite{S} Scheunert has introduced the concept of $\ep$-algebra, which seems to be the widest generalization of super-algebra. The sign appearing in super-algebra is generalized in $\ep$-algebra by a so-called ``commutation factor''. We will define this concept in section 2, and quickly review some basic facts, most of which can be found in \cite{S} or \cite{KN1}. However, we will emphasize on free $\ep$-modules, since they do not behave as simply as super-modules, something that is not always clearly stated. We will define three different notions of rank for these modules and give a sufficient condition for an $\ep$-algebra to have invariant basis number, in the sense of Cohn (see \cite{C}).

Although the existence of $\ep$-algebras is quite well known, their study has not been undertaken with the same energy as the study of super-algebras mainly for two reasons : they are much too general and they have not proved to be useful for physics yet. Nevertheless, the particular case where the gradation is over $\ZZ$ and $\ep=(-1)^d$, with $d$ an (anti-)symmetric bilinear form seems to be the first logical step away from super-symmetry if ever this step has to be made. This is precisely this kind of $\ep$-algebras we were naturally led to while studying ``number operator algebras''. The latter are algebras we expect to play a role in the quantization procedure of the equations of motion of a system of harmonic oscillators (see \cite{Bes}). We define them and give a classification theorem in section 3.

Interestingly enough, every number operator algebra depends on a single constant $h$ (which plays the role of the Planck constant). It is then natural to try to interpret the algebras at $h\not=0$ (quantum algebras) as deformations of those at $h=0$ (classical algebras) as we shall see in section 4.

In section 5, we show that except in two special cases that happen only for a finite number of degrees of freedom (thus, not in field theory) and which deserve to be studied on their own, the classical algebras are endowed with an $\ep$-Poisson structure coming from the $\ep$-Lie structure on the quantum algebras. This $\ep$-Poisson structure can be used to formulate ``classical'' equations of motion for the system of harmonic oscillators under consideration.

Throughout the article, we denote by $K$ a field, and by $K\langle X\rangle$ the free algebra generated by a set $X$ over $K$. All rings and algebras are unital, and morphisms preserve unit.

\section{$\ep$-algebra}
\subsection{Basic definitions}
The following definition comes from \cite{Bou} (chap. III, p. 46. See also p. 116). Note that, in contrast with \cite{S} and \cite{KN1}, we do not assume that $G$ is a group.

\begin{definition}
Let $(G,+)$ be a commutative monoid. A commutation factor on $G$ with values in $K$ is a mapping $\ep : G\times G\rightarrow K$ such that :
\bea
\ep(g,h)\ep(h,g)=1 \label{c1}\\ 
\ep(g+h,k)=\ep(g,k)\ep(h,k)\label{c2}
\eea
\end{definition}
>From (\ref{c1}) and (\ref{c2}) one sees at once that $\ep(g,h+k)=\ep(g,h)\ep(g,k)$. Of course, the range of $\ep$ is included in $K^\times$, the group of invertible elements of $K$. Moreover, from (\ref{c1}), we have $\ep(g,g)^2=1$ for any $g\in G$. >From (\ref{c2}) we also have $\ep(0,g)=\ep(g,0)=1$. In particular $\ep(0,0)=1$. If $g$ has an opposite $g'$, we see from (\ref{c2}) that $\ep(g',k)=\ep(g,k)^{-1}$. Thus, if $G$ is a semi-group, we can extend $\ep$ uniquely to the group obtained from $G$ by symmetrization. It is obvious that this extension is a commutation factor.

We now define a map $p$ (for ``parity''), from $G$ to $\ZZ/2\ZZ$ by $p(g)=0$ if $\ep(g,g)=1$ and $p(g)=1$ if $\ep(g,g)=-1$. Since $\ep(g+h,g+h)=\ep(g,g)\ep(g,h)\ep(h,g)\ep(h,h)$ $=\ep(g,g)\ep(h,h)$, and $\ep(0,0)=1$, $p$ is a monoid homomorphism from $(G,+)$ to $(\ZZ/2\ZZ,+)$. Let us call $G_0=Ker(p)$, and $G_1=G\setminus G_0$. The elements in $G_0$ are called ``even'', the ones in $G_1$ are called ``odd''.

Note that in order to have $G_1\not=\emptyset$ we must impose a condition on $G$. Indeed, since $\ep(n.g,g)=\ep(g,g)^n$ every odd element must be of even order. For instance, if $G=\ZZ/p\ZZ$ with $p$ odd then $G_1=\emptyset$ for every $\epsilon$ defined on $G$.

\smallbreak
Now for an example : let $G=\ZZ\times\ZZ$, take $q\in K^\times$, and set $\ep_q((k,l),(m,n))=q^{lm-kn}$. More generally, for any $G$ consider a monoid homomorphism $\phi : G\times G\rightarrow(\ZZ,+)$ which is anti-symmetric, then for any $q\in K^\times$, $\ep=q^\phi$ is a commutation factor. Note that if $q=\pm 1$, $\phi$ may either be taken symmetric or anti-symmetric.
 
\begin{definition}
Let $A$ be a $K$-algebra. $A$ is called a $G$-graded $K$-algebra iff there exist $K$-subspaces $(A_g)_{g\in G}$, such that :
\bea
A=\bigoplus_{g\in G}A_g\\
A_gA_h\subset A_{g+h}
\eea
\end{definition}

An element in $A_g$ for some $g$ is called homogenous. In the sequel, we use the following notation : if $a\in A_g$, we write $\bar a=g$. More generally, if $x$ is a homogenous element in any $G$-graded object, we will write $\bar x$ for the grade of $x$. 

We also introduce the notation ``$\forall_h$'' that will mean ``for all homogenous''. For example, $\forall_h x\in A$ means ``for all homogenous $x$ in $A$''.

A couple $(A,\ep)$ where $A$ is a $G$-graded $K$-algebra, and $\ep$ is a commutation factor on $G$ with values in $K$ is called an ``$\ep$-algebra''. We will also say ``$A$ is an $\ep$-algebra''. 

If $G$ is a semi-group, it can be embedded into a group $G'$. Since $\ep$ extends to a commutation factor $\ep'$ on $G'$, any $\ep$-algebra can be considered as an $\ep'$-algebra with $A_g=\{0\}$ for $g\in G'\setminus G$. Thus, we can always work with groups instead of semi-groups.

Let us look at an example. Let $M_q$ be Manin's quantum plane, that is : $M_q=K\langle x,y\rangle/\langle xy-qyx\rangle$, where $\langle xy-qyx\rangle$ is the two-sided ideal generated by $xy-qyx$, with $q\in K^\times$. Since $\{y^kx^l|k,l\in\NN\}$ is a $K$-basis of $M_q$, one sees at once that there is a unique $\NN\times\NN$-grading
 such that $\bar x=(1,0)$, $\bar y=(0,1)$, endowing $(M_q,\ep_q)$ with the structure of an $\ep_q$-algebra. 
\smallbreak
Throughout the rest of this section, $A$ will be a fixed $\ep$-algebra.

$\forall_h$ $x,y\in A$, we define the $\ep$-commutator by : $$[x,y]_\ep=xy-\ep(\bar x,\bar y)yx$$
It is then extended to non-homogenous elements by linearity. We say that $x$ and $y$ $\ep$-commute iff $[x,y]_\ep=0$. The $\ep$-center of $A$, $Z_\ep(A)$, is the set $\{x\in A|\forall y\in A, [x,y]_\ep=0\}$. $A$ is said to be $\ep$-commutative iff $Z_\ep(A)=A$. For instance, $M_q$ is $\ep_q$-commutative.

There is a super-algebra $\tilde A$ naturally associated with $A$ : it is defined by $\tilde A_0=\bigoplus_{g\in G_0}A_g$, and $\tilde A_1=\bigoplus_{g\in G_1}A_g$. If $A$ is $\ep$-commutative, it is not true in general that $\tilde A$ is super-commutative. However, it is easy to see that in an $\ep$-commutative algebra, every $x\in\tilde A_1$ is nilpotent.

In the next two definitions, we assume the characteristic of $K$ to be $\not=2,3$.
\begin{definition}
Let $V$ be a $G$-graded $K$-space, and $[.,.]$ be a bilinear map from $V\times V$ to $V$, such that $\forall_h$ $x,y,z\in V$ :
\be
[x,y]=-\ep(\bar y,\bar x)[y,x]
\ee
\be
\ep(\bar z,\bar x)[x,[y,z]]+\ep(\bar y,\bar z)[z,[x,y]]+\ep(\bar x,\bar y)[y,[z,x]]=0
\ee
$V$ is called an $\ep$-Lie algebra.
\end{definition} 

For example, $(A,[.,.]_\ep)$ is an $\ep$-Lie algebra.

\begin{definition}
Let $\{.,.\} : A\times A\rightarrow A$ be a $K$-bilinear map. $\{.,.\}$ is called an $\ep$-Poisson bracket, and $A$ is an $\ep$-Poisson algebra, if and only if :
\begin{enumerate}
\item $(A,\{.,.\})$ is an $\ep$-Lie algebra
\item $\forall_h\ x,y,z\in A,\ \{x,yz\}=\{x,y\}z+\ep(\bar x,\bar y)y\{x,z\}$
\end{enumerate}
\end{definition}

\subsection{$\ep$-modules}

They are just $G$-graded $A$-modules. More precisely, a left module $M$ over $A$ is called a left $\ep$-module over $A$ if, and only if, there exists a decomposition $M=\bigoplus_{g\in G}M_g$ as $K$-space, such that $A_gM_h\subset M_{g+h}$. Right $\ep$-modules are similarly defined.

Given an $\ep$-module over $A$, a super-module $\tilde M$ over $\tilde A$ is defined by $\tilde M_0=\bigoplus_{g\in G_0}M_g$ and $\tilde M_1=\bigoplus_{g\in G_1}M_g$.

If $A$ is $\ep$-commutative, there exists on any left $\ep$-module a canonical right $\ep$-module structure compatible with it, given by :
\be
\forall_h a\in A, \forall_h m\in M,\quad m.a=\ep(\bar m,\bar a)a.m\label{rs}
\ee

Let $M$, $N$ be two left $\ep$-modules over $A$. For all $\gamma\in G$, we define a homomorphism of grade $\gamma$ from $M$ to $N$ to be a $K$-linear map $f$ such that\footnote{if $G$ is not a semi-group, the same homomorphism may have several grades}
\be
\forall g\in G,\quad f(M_g)\subset N_{g+\gamma}\label{h1}
\ee
and furthermore :
\be
\forall_h a\in A, \forall_h m\in M,\quad f(a.m)=\ep(\gamma,\bar a)f(m)\label{h2}
\ee
We denote by  $\Hom^\gamma_A(M,N)$ the set of those morphisms. For two right modules, a homomorphism of grade $\gamma$ has to fulfill (\ref{h1}), but (\ref{h2}) is replaced by
\be
f(m.a)=f(m).a\label{h3}
\ee
One sees that with these definitions, if $A$ is $\ep$-commutative, any grade $\gamma$ left homomorphism is automatically a grade $\gamma$ right homomorphism for the right structure (\ref{rs}).

We will be mainly concerned with grade 0 homomorphisms over an $\ep$-commuta\-tive $A$.

A free $\ep$-module is just a free graded module, that is a graded module in which a homogenous basis exists. Note that if $M$ is free as an ungraded module, it need not be free as a graded module, even if $A$ is commutative. An important exception is when $M$ is a super-module over a super-commutative algebra. In this case, given an ungraded basis one can find a graded one (see \cite{D} p 19).

We will be only concerned with finite free $\ep$-modules. There are three natural notions of rank for them.

Let us recall that a ring $R$ is said to have ``invariant basis number'' (IBN) if (see \cite{C}) : $$R^m\simeq R^n\Rightarrow m=n$$

It is equivalent to saying that for any two matrices $P\in{\cal M}_{m,n}(R)$ and $Q\in{\cal M}_{n,m}(R)$ : $$(PQ=I_m \mbox{ and } QP=I_n)\Rightarrow m=n$$ If there exists a ring homomorphism from $R$ to $S$, we can extend it to a ring homomorphism of matrix algebras, thus $S$ has IBN $\Rightarrow$ $R$ has IBN.
\smallbreak
 
Let ${\bf n} : G\rightarrow\NN$ be a map such that the set $\mbox{Supp}({\bf n})=\{g\in G|{\bf n}(g)\not=0\}$ is finite. We define the (right, say) module $F=A^{\bf n}$ to be the direct sum $\bigoplus_{g\in \mbox{Supp}({\bf n})}A^{{\bf n}(g)}$. If $\{e_g^i|g\in \mbox{Supp}({\bf n}), 1\leq i\leq {\bf n}(g)\}$ is the canonical basis, we define a grading by $\bar e_g^i=g$.

\begin{definition}
The integer $n=\sum_g{\bf n}(g)$ is called the ``total rank'' of $F$.

The couple $(p,q)$ where $p=\sum_{g\in G_0} {\bf n}(g)$, $q=\sum_{g\in G_1}{\bf n}(g)$, denoted by $p|q$, is called the ``super-rank'' of $F$. ${\bf n}$ is called the ``$\ep$-rank'' of $F$. 
\end{definition}

$F$ is the canonical (right) free $\ep$-module over $A$ of $\ep$-rank ${\bf n}$. If two canonical free $\ep$-modules over $A$ that are isomorphic must have the same $\ep$-rank, we say that $A$ has invariant $\ep$-rank, and the $\ep$-rank is uniquely defined for any free $\ep$-module over $A$. The properties of having ``invariant super-rank'' and ``invariant total rank'' are similarly defined. Of course $A$ has invariant $\ep$-rank $\Rightarrow$ $A$ has invariant super-rank $\Rightarrow$ $A$ has invariant total rank.


Elements of $\Hom^\gamma(A^{\bf m},A^{\bf n})$ may be represented by matrices in ${\cal M}_{n,m}(A)$. For more details see \cite{KN1} (but note that their matrices must act on the right of row vectors, which represent elements of a left module, whereas if we prefer working with right modules, we must take column vectors, with matrices acting on the left, and this is what we will do).

\begin{theorem}\label{ibn}
Let $A$ be an $\ep$-algebra and $B$ be an algebra having IBN. Denote by $H$ the set of algebra homomorphisms from $A$ to $B$. Then :
\begin{enumerate}
\item  If $H\not=\emptyset$, $A$ has invariant total rank.
\item  If $\exists\pi$, $\pi\in H$, $\pi(A_g)=\{0\}$ for all $g\in G_1$, then $A$ has invariant super-rank.
\item If $\exists\pi$, $\pi\in H$, $\pi(A_g)=\{0\}$ for all $g\not=0$, then $A$ has invariant $\ep$-rank
\end{enumerate}
\end{theorem}
\dem\newline
The first assertion comes from the trivial fact that a homogenous basis is also an ungraded basis.
\smallbreak
For the second assertion we take 2 homogenous bases $\{e_\rho,u_\sigma|1\leq\rho\leq m, 1\leq\sigma\leq k\}$ and $\{f_\alpha,v_\beta|1\leq\alpha\leq n, 1\leq\beta\leq l\}$ of a free $\ep$-module over $A$, such that the $e_\rho$ and the $f_\alpha$ are even, and the $u_\sigma$, $v_\beta$ are odd.    
We write : $f_\alpha=\sum_\rho e_\rho a_{\rho,\alpha}+\sum_\sigma u_\sigma b_{\sigma,\alpha}$, and $v_\beta=\sum_\rho e_\rho c_{\rho,\beta}+\sum_\sigma u_\sigma d_{\sigma,\beta}$.

Of course we also have : $e_\rho=\sum_\mu f_\mu a'_{\mu,\rho}+\sum_\nu v_\nu b'_{\nu,\rho}$, and $u_\sigma=\sum_\mu f_\mu c'_{\mu,\sigma}+\sum_\nu v_\nu d'_{\nu,\sigma}$.

We can suppose without loss of generality that :
$$a_{\rho,\alpha}\in\bigoplus_{g|g+\bar e_\rho=\bar f_\alpha} A^g$$
and
$$b_{\sigma,\alpha}\in\bigoplus_{g|g+\bar u_\sigma=\bar f_\alpha}A^g$$
and similarly for the other coefficients. We have the matrix equalities :
\be
\matrix{ &\matrix{ {\scriptstyle m} &{\scriptstyle k}}\cr
\matrix{{\scriptstyle n}\cr {\scriptstyle l}}&\left(\matrix{a'&c'\cr b'&d'}\right)}\matrix{ &\matrix{ {\scriptstyle n} &{\scriptstyle l}}\cr
\matrix{{ }_m\cr { }_k}&\left(\matrix{a&c\cr b&d}\right)}=\matrix{ &\matrix{ & }\cr
\matrix{ \cr }&\left(\matrix{I_n&0\cr 0&I_l}\right)}\label{mat1}
\ee
and
\be
\matrix{ &\matrix{ {\scriptstyle n} &{\scriptstyle l}}\cr
\matrix{{ }_m\cr { }_k}&\left(\matrix{a&c\cr b&d}\right)}\matrix{ &\matrix{ {\scriptstyle m} &{\scriptstyle k}}\cr
\matrix{{\scriptstyle n}\cr {\scriptstyle l}}&\left(\matrix{a'&c'\cr b'&d'}\right)}=\matrix{ &\matrix{ & }\cr
\matrix{ \cr }&\left(\matrix{I_m&0\cr 0&I_k}\right)}\label{mat2}
\ee
Since the parity map is a monoid homomorphism, the coefficients of $b$, $b'$, $c$, $c'$ are odd. Thus the images under $\pi$ of (\ref{mat1}) and (\ref{mat2}) are :
\be
\left(\matrix{\pi(a')&0\cr 0&\pi(d')}\right)\left(\matrix{\pi(a)&0\cr 0&\pi(d)}\right)=\left(\matrix{I_n&0\cr 0&I_l}\right)
\ee
and
\be
\left(\matrix{\pi(a)&0\cr 0&\pi(d)}\right)\left(\matrix{\pi(a')&0\cr 0&\pi(d')}\right)=\left(\matrix{I_m&0\cr 0&I_k}\right)
\ee
Therefore $\pi(a')\pi(a)=I_n$ and $\pi(a)\pi(a')=I_m$. Since $B$ has IBN, we get $m=n$. In the same way, we find $k=l$.
\smallbreak
For the last assertion, we take two bases of $\ep$-rank ${\bf m}$ and ${\bf n}$. Let us suppose first that $G$ is a group.

As before, the change of bases gives the following matrix equations 
\be
\left(\matrix{ & & \cr  & a_{g,h} & \cr & & }\right)\left(\matrix{ & & \cr  & a'_{g,h} & \cr & & }\right)=I_m
\ee
\be
\left(\matrix{ & & \cr  & a'_{g,h} & \cr & & }\right)\left(\matrix{ & & \cr  & a_{g,h} & \cr & & }\right)=I_n
\ee
where the $a_{g,h}$ and $a'_{g,h}$ are block matrices. They are indexed by the elements of $\mbox{Supp}({\bf m})\cup\mbox{Supp}({\bf n})$. Their coefficients are of grade $g-h$. The block $a_{g,h}$ has ${\bf m}(g)$ rows and ${\bf n}(h)$ columns, and $a'_{g,h}$ has ${\bf n}(g)$ rows and ${\bf m}(h)$ columns. $I_m$ and $I_n$ are the identity matrices of rank $m=\sum_g{\bf m}(g)$ and $n=\sum_g{\bf n}(g)$.

Taking the image under $\pi$ makes all the off-diagonal block matrices vanish. We thus find for all $g\in\mbox{Supp}({\bf m})\cup\mbox{Supp}({\bf n})$ : $\pi(a_{g,g})\pi(a'_{g,g})=I_{{\bf m}(g)}$ and $\pi(a'_{g,g})\pi(a_{g,g})=I_{{\bf n}(g)}$. Since $B$ has IBN, we see that ${\bf m}$ and ${\bf n}$ must coincide on their support. Thus, they are equal.

In the general case the coefficients of the matrices $a_{g,h}$ and $a_{g,h}'$ belong to $\bigoplus_{k\in G|k+h=g}A^k$, and if $g\not=h$, we have $k\not=0$, and $\pi(A^k)=0$. The result follows.

\hfill QED.

\smallbreak
It is well known that commutative algebras have invariant total rank, and that super-commutative algebras have invariant super-rank (cf. \cite{D}). But the following example shows that the invariance of the $\ep$-rank is not always true, even for an $\ep$-commutative $A$.

Consider the algebra $A=K\{x,y,x^{-1},y^{-1}\}$ generated by variables $x$,\ldots,$y^{-1}$ such that $x$ and $x^{-1}$ anti-commute with $y$ and $y^{-1}$. A $K$-basis for $A$ is given by $\{x^ky^l|k,l\in\ZZ\}$, a $\ZZ/2\ZZ\times\ZZ/2\ZZ$-grading is uniquely defined by $\bar x=(1,0)$, $\bar y=(0,1)$. Finally, $A$ is $\ep$-commutative with $\ep((k,l),(m,n))=(-1)^{kn+lm}$. But $x$ and $y$ are two homogenous bases of $A$ considered as a module over itself, with different $\ep$-rank.
\smallbreak
Nevertheless, there are many cases in which the conditions of theorem \ref{ibn} are met. For instance, an $\ep$-commutative $A$ can sometimes be written $A=C\oplus I$, where $I$ is a two-sided ideal, and $C$ is a commutative sub-algebra. Then of course the hypothesis of theorem \ref{ibn} is satisfied by the quotient map. An example of this situation is when $G=\NN^k$, or more generally, the semi-group of positive elements of some ordered group. Then we can take $C=A_0$ and $I=\bigoplus_{g\not=0}A_g$. Another situation that we shall meet is when $C=K.1$, in which case $A$ can be viewed as a non-unital algebra to which a unit has been added.

\subsection{Tensor products}

In this paragraph, $A$ is $\ep$-commutative, all modules are considered as $A-A$-bimodules through (\ref{rs}), and all bases are homogenous.

Let $V_1,\ldots,V_n$ be $\ep$-modules over $A$. Their tensor product $V_1\otimes\ldots\otimes V_n$ is defined using the general construction given in \cite{Bou} (III, p. 65-69). In particular, note that for $a\in A$, $v_i\in V_i$, one has :
$$v_1\otimes\ldots\otimes v_i.a\otimes v_{i+1}\otimes\ldots v_n=v_1\otimes\ldots\otimes v_i\otimes a.v_{i+1}\otimes\ldots\otimes v_n$$
Note also that the commutativity isomorphism $V_1\otimes V_2\simeq V_2\otimes V_1$ should be defined by $v_1\otimes v_2\mapsto \ep(\bar v_1,\bar v_2)v_2\otimes v_1$.

We then define the $\ep$-tensor algebra of $V$ : $T_\ep(V)=\bigoplus_{n\in\NN}V^{\otimes n}$. The gradation $\overline{v_1\otimes\ldots\otimes v_n}=\bar v_1+\ldots+\bar v_n$ turns it into an $\ep$-algebra.

In particular, if $V$ is a free $\ep$-module with basis $\{x_1,\ldots,x_n\}$, $A_\ep\langle x_1,\ldots,x_n\rangle:=T_\ep(V)$ is the free $\ep$-algebra over $A$ on the generators $\{x_1,\ldots,x_n\}$.

In general, all usual constructions carry over, provided one puts in an epsilon term each time two factors are exchanged. For instance, one can define the $\ep$-antisymmetric algebra $\Lambda_\ep(V)=T_\ep(V)/I$, where $I$ is the two-sided ideal generated by the elements of the form $v\otimes w+\ep(\bar v,\bar w)w\otimes v$. Note that if any generator is odd this algebra is of infinite rank, thus we cannot use it to define a determinant. If we had chosen not to put the epsilon factor in the definition of $\Lambda_\ep(V)$, we would have had an algebra of finite rank with the top exterior product of rank one, and with basis $x_1\wedge\ldots\wedge x_n$, as usual. However, this construction is not functorial when there are odd elements, so the usual determinant cannot be defined. Nevertheless, the generalization of the determinant to super-algebra, known as the Berezinian, can be extended to $\epsilon$-algebra, as it is shown in \cite{KN1}.

\underline{Remark} : For the reader acquainted with these matters, we mention here that Bergman's diamond lemma, which is a most useful result, is valid in this setting for graded reduction systems, that is to say systems $\{(w_\sigma,f_\sigma)|\sigma\in S\}$ where $f_\sigma$ is homogenous of grade $\bar w_\sigma$. This is an immediate application of \cite{Ber}, section 6.

\section{Number Operator Algebras}
In \cite{Bes} we have introduced number operator algebras.

\begin{definition}
Let $K$ be a field of characteristic 0 endowed with an involutive automorphism $\tau$, $B$ a non-trivial $K$-algebra (that is $B\not=0$, $B\not=K$), $Z(B)$ the centre of $B$, and let $C^+=\{\aix|i\in\II\}$, $C^-=\{\ai|i\in\II\}$, and $N=\{N_i|i\in\II\}$ be 3 sets indexed by $\II$, with the $\ai$'s and $\aix$'s in $B$, and $N_i$'s in $B/Z(B)$. $(B,C^+,C^-,N)$ is said to be a number operator algebra if, and only if :

\item{(i)} $B$ is generated by $C^+\cup C^-$.
\item{(ii)} One uniquely defines an anti-involution $J$ on $B$ by setting $J(\ai)=\aix$.
\item{(iii)} $\forall i,j\in\II$, $[N_i,\ajx]=\delta_{ij}\ajx$ and $[N_i,\aj]=-\delta_{ij}\aj$.
\end{definition}

\underline{Remarks} :
\begin{itemize}
\item{}In the physical case $K=\CC$ and $\tau$ is the complex conjugation.
\item{}We shall say ``$B$ is a n.o.a.'' rather than using the lengthy expression ``$(B,C^+,C^-,N)$ is a number operator algebra''.
\item{}We say that $B$ is of type $\alpha$ when $\alpha$ is the cardinal of $\II$.
\end{itemize}
We go on with a few more definitions. From now on we fix a n.o.a. $B$, with all its features, $C^+$, $C^-$, etc\ldots

Let us call $L$ the free $K$-algebra generated by $C^+\cup C^-$, and $\pi$ the canonical morphism from $L$ onto $B$. If the kernel of $\pi$ is generated by elements of degree two or less, we say that $B$ is quadratically presented.

Let $\SS_\II$ be the group of one-one mappings from $\II$ to $\II$ leaving all elements invariant except for a finite number. Every $\sigma$ in $\SS_\II$ naturally gives rise to an algebra automorphism of $L$, denoted by $\sigma^*$, such that for all $i\in\II$, $\sigma^*(\ai)=a_{\sigma(i)}^{}$, and $\sigma^*(\aix)=a_{\sigma(i)}^+$.

If every such $\sigma^*$ induces an automorphism of $B$, we say that $B$ is symmetric.

It is possible to classify all n.o.a. of infinite type which are symmetric and quadratically presented (see \cite{Bes}).

\begin{theorem}\label{clas}
Let $B$ be symmetric and quadratically presented. If $\II$ is infinite, then there exists an $h\in K^+\setminus\{0\}$ ($K^+$ is the sub-field of elements of $K$ that are invariant under $\tau$) such that Ker$(\pi)$ is generated by one of the following sets :
\item{(a)} $\{\ai^2,\aix^2,\ai\aj+\aj\ai,\aix\ajx+\ajx\aix,\ai\ajx+\ajx\ai,\ai\aix+\aix\ai-h|i,j\in\II,i\not=j\}$ (Fermionic case)
\item{(a')} $\{\ai^2,\aix^2,\ai\aj-\aj\ai,\aix\ajx-\ajx\aix,\ai\ajx-\ajx\ai,\ai\aix+\aix\ai-h|i,j\in\II,i\not=j\}$ (Pseudo-Fermionic case)
\item{(c)} $\{\ai\aj-\aj\ai,\aix\ajx-\ajx\aix,\ai\ajx-\ajx\ai,\ai\aix-\aix\ai-h|i,j\in\II, i\not=j\}$ (Bosonic case)
\item{(c')} $\{\ai\aj+\aj\ai,\aix\ajx+\ajx\aix,\ai\ajx+\ajx\ai,\ai\aix-\aix\ai-h|i,j\in\II, i\not=j\}$ (Pseudo-Bosonic case)
\end{theorem}

We denote by $\hat \fer^h$, $\fer^h$, $\bos^h$ and $\hat\bos^h$, respectively, the algebras corresponding to each of these four cases. Actually we should call them $\hat\fer^{h,\II}$, etc\ldots but we assume that $\II$ is fixed and thus no confusion can be made.

It should be noted that although these four kinds of algebras are generally not isomorphic to each other, this can still happen for $\hat\fer^h$ and $\fer^h$, at least when $\II$ is countable (Brauer-Weyl isomorphism).

Nevertheless, these four cases are distinct as symmetric n.o.a. : that is to say, there exists no isomorphism $\phi$ between any two of them, such that $\phi$ commutes with the anti-involution and with the action of $\SS_\II$, and such that $\phi$ send any number operator to a number operator. Of course, all this can be formulated in terms of categories.

What is the situation inside each of the four cases ?
We see that the algebras only depend on a parameter $h\in K^+$. Consider $B^h$ and $B^{h'}$, two n.o.a. of one of the four species (either two algebras of fermions, or two algebras of pseudo-fermions, or etc\ldots). Then one can show that these algebras are isomorphic as symmetric number operator algebras iff $\exists\lambda\in K^+$ such that $h'=\lambda\tau(\lambda)h$. When such a relation exists between $h$ and $h'$, let us define the mapping $\phi_\lambda$, given by $\phi_\lambda(\ai)=\lambda\ai'$, $\phi_\lambda(\aix)=\tau(\lambda)\aix'$, where the elements with a prime are in $B^{h'}$ and the others are in $B^h$. $\phi_\lambda$ is easily found to have all the required properties for an isomorphism of symmetric number operator algebras.

In the physical case, we see that the four families of theorem 2 depend on a non-zero real constant, and furthermore in each of the four families there are exactly two isomorphism classes in the category of symmetric n.o.a., one for $h>0$ and the other for $h<0$, which makes 8 isomorphism classes as a whole. The isomorphisms $\phi_\lambda$ clearly correspond to a rescaling of the units. 

\underline{Remark} : When $\alpha$ is finite, a classification can be done under a supplementary hypothesis, namely the ``confluence hypothesis''. In this case we have two more families of algebras, which have the following presentation : 
\begin{itemize}
\item{(b)} $\{\ai^2,\aix^2,\ai\aj,\aix\ajx,\ai\ajx,\ai\aix+\sum_{k\in\II}\akx\ak-h|i,j\in\II,i\not=j\}$
\item{(b')} $\{\ai^2,\aix^2,\ai\aj,\aix\ajx,\ajx\ai,\aix\ai+\sum_{k\in\II}\ak\akx-h|i,j\in\II,i\not=j\}$
\end{itemize}

Of course, these two kinds of algebras are isomorphic as algebras (and are isomorphic to matrix algebras), but once again not as number operator algebras (except when $n=1$, in which case only two kinds of algebras remain\penalty 1000 : bosonic and fermionic). We call them ${\cal E}^h$ and ${\cal E'}^{h}$. In the physical case, there are also two isomorphism classes in each case $(b)$ or $(b')$, distinguished by the sign of $h$, in the category of number operator algebras. Particles corresponding to such algebras would follow an exclusion principle even more severe than Pauli's : only one such particle could exist at a given time, regardless of its state.

\section{Classical limit of Number Operator Algebras}

We quickly recall the definition of formal deformations of algebras. See \cite{Ger} or \cite{Gui} for more details.

Let $B$ be a $K$-algebra whose multiplication is seen as a bilinear map $\mu : B\times B\rightarrow B$. Let $K[[\hbar]]$ stand for the algebra of formal series in $\hbar$, and $\tilde B$ stand for the $K[[\hbar]]$-algebra of formal series with coefficients in $B$.

A ``formal deformation of $(B,\mu)$'' is a $K[[\hbar]]$-algebra structure $\tilde\mu$ on $\tilde B$, such that the canonical map $\tilde B/\hbar\tilde B\rightarrow B$ is an algebra isomorphism. These data are equivalent to the existence of a sequence of bilinear maps $\mu_n: B\times B\rightarrow B$, with $\mu_0=\mu$, such that :
\be
\forall x,y,z\in B,\, \forall n\geq 1,\, \sum_{p+q=n}(\mu_p(\mu_q(x,y),z)-\mu_p(x,\mu_q(y,z)))=0\label{st}
\ee

$\tilde\mu$ is then defined by setting, for all $x,y\in B\subset\tilde B$ :
\be
\tilde\mu(x,y):=\sum_n\mu_n(x,y)\hbar^n\label{stst}
\ee

and extending to formal series in the obvious way.

It is sometimes possible to ``fix the parameter'', i.e. to replace everywhere $\hbar$ by a constant $h\in K$. In particular, this is the case when the right-hand side of (\ref{stst}) is a polynomial for every $x,y\in B$. By doing so, a new algebra structure $\mu^h$ is defined on $B$. We shall say that $(B,\mu^h)$ is a deformation of $(B,\mu)$.

Let $B^h$ be a n.o.a. If we replace $h$ by $0$ in the presentation of $B^h$, we get a new algebra $B^0$ that we call the ``classical limit'' of $B^h$. Let us see what these algebras look like in the cases $(a)$, $(a')$, $(c)$ and $(c')$ of theorem \ref{clas}.

It is easy to see that $\hat\fer^0$ is the exterior algebra $\Lambda[\ai,\aix|i\in\II]$ over variables $\ai$ and $\aix$, and that ${\cal A}^0$ is the symmetric (or polynomial) algebra $S[\ai,\aix|i\in\II]$ over the same variables. The other two have no names (however $\fer^0$ is a particular case of the generalized Grassman algebras of \cite{OK}), but we have the following algebra isomorphisms :

$$\fer^0\simeq\tens_\II\Lambda[a,a^+]\qquad\hat \bos^0\simeq\hattens_\II S[a,a^+]$$
where $\hattens$ means ``graded tensor product''. 

Let us have a closer look at $\bos^h$ and $\bos^0$. Fix a total ordering $<$ on $\II$. Take two tuples of indices $\JJ=(j_1,\ldots,j_r)$ and $\KK=(k_1,\ldots,k_s)$ such that $j_1\leq\ldots\leq j_r$, $k_1\leq\ldots\leq k_s$, and set $\ajjx:=a_{j_1}^+\ldots a_{j_r}^+$ and $\akk:=a_{k_1}^{}\ldots a_{k_s}^{}$. If $\JJ$ or $\KK$ is empty we set $a_\emptyset^{}=a_\emptyset^+=1$. ${\cal A}^h$ has a basis $T$ of the form : $\{\ajjx\akk|\JJ$ and $\KK$ are any ordered tuples of indices$\}$. But $\bos^0$ has a basis of the same form and we can use it to identify $\bos^0$ and $\bos^h$ as vector spaces. Now take any two elements of $T$, multiply them in $\bos^h$, and write the result in terms of basis vectors : we can see it as a polynomial in $h$ and take the coefficients to define $\mu_n$ as in the formula below :
$$\mu(x,y)=:\sum_{n\geq 0}\mu_n(x,y)h^n$$
where $\mu$ is the multiplication of $\bos^h$.

We can then extend $\mu_n$ by bilinearity. We claim that the $\mu_n$ just defined fulfil the conditions (\ref{st}) and that $\mu_0$ is actually the multiplication of $\bos^0$. To show it we should use the concepts of reduction systems and confluence (see \cite{Ber}). The reader acquainted with these matters will see it to be an easy consequence of the confluence of the reduction system $S=\{(\ai\aj,\aj\ai),(\aix\ajx,\ajx\aix),(\ai\ajx,\ajx\ai),(\aj\aix,\aix\aj),(\ai\aix,\aix\ai+h)|i,\penalty -100 j\in\II,$ $i<j\}$ for any value of $h$. We refer to \cite{Gui} (Chap. 3, theorem 2.6.2.) for more details.

We can thus define a formal deformation $\tilde\mu$ of $\bos^0$, and if we set the constant to $h$, we obviously have $(\bos^0,\mu^h)\simeq (\bos^h,\mu)$.

We can do the same with $\hat\bos^h$, $\hat\fer^h$ and $\fer^h$.

\underline{Remark} : Let us say a little word about the case $(b)$ ($(b')$ being symmetrical). ${\cal E}^h$ is a deformation of ${\cal E}^0$, which is a $(\alpha+1)^2$-dimensional local algebra. We do not know if ${\cal E}^0$ has ever been considered. Since it is not an $\ep$-Poisson algebra, we will not study it in the next section. Nevertheless, it has an interesting structure that we plan to study in another article.

\section{$\ep$-Poisson structures}
Let us begin by the example of $\bos^h$, which is well known.

Since $\bos^0$ is commutative, the formula 
$$\{x,y\}:=\mu_1(x,y)-\mu_1(y,x)$$
defines a Poisson bracket on $\bos^0$ (see \cite{Gui})\footnote{To follow usual conventions, one should divide out by $i$ in the physical case}. It can also be written in the heuristic form :
$$\{x,y\}=\lim_{\hbar\rightarrow 0}{1\over\hbar}(\tilde\mu(x,y)-\tilde\mu(y,x))$$
This last formula is useful to see that all the properties of the Poisson bracket come from the corresponding properties of the normal (commutator) bracket of $\tilde\mu$, to the first order in $\hbar$. $\bos^0$ is thus a Poisson algebra.
\smallbreak

$\hat\fer^h$, $\fer^h$ $\bos^h$ and $\hat\bos^h$, as well as their classical limits are naturally $\ZZ^{(\II)}$-graded (where $(\II)$ means direct sum over $\II$). It comes from the gradation on the free algebra $L$ uniquely defined by $\bar\ai^+=p_i$, $\bar\ai=-p_i$, where $p_i$ is the element of $\ZZ^{(\II)}$ defined by $p_i(j)=\delta_{ij}$.

We call it the gradation by the number of particles. Since the ideals of definition of quantum as well as classical algebras are homogenous with respect to this gradation, it goes to the quotient in both cases.

Let us define bilinear maps from $\ZZ^{(\II)}\times\ZZ^{(\II)}$ to $\ZZ$ :
$$d_{a}(p,q):=(\sum_{i\in\II}p(i))(\sum_{i\in\II}q(i))$$
$$d_{a'}(p,q):=\sum_{i\in\II}p(i)q(i)$$
$$d_c(p,q):=0$$
$$d_{c'}(p,q)=\sum_{i,j\in\II \atop i\not=j}p(i)q(j)$$
We then define commutation factors on $\ZZ^{(\II)}$ by $\ep_l(p,q)=(-1)^{d_l(p,q)}$, with $l=a$, $a'$, $c$ or $c'$.

\begin{theorem}
$\hat\fer^0$ is an $\ep_a$-commutative $\ep_a$-algebra, $\fer^0$ is $\ep_{a'}$-commutative, $\bos^0$ is $\ep_c$-commutative (that is to say commutative), and $\hat\bos^0$ is $\ep_{c'}$-commutative.

Moreover, they have invariant $\ep$-rank.
\end{theorem}
\dem\newline
Let us examine for instance the case of $\hat\bos^0$ : take a basis element $\aiix\ajj$, $p$ its grading, and $\ak$ a generator. We see that $\ak$ anti-commutes with everything, except with $\ak$ and $\akx$, to which it commutes. Thus we find $\aiix\ajj\ak=(-1)^{\sum_{i\not=k}p_i}\ak\aiix\ajj$. It is of course the same for $\akx$, and by iteration, we find the commutation rules for two basis elements. Then it is easy to extend the result to homogenous elements.

Now the map $\pi : \hat\bos^0\rightarrow K$ given by the projection on the basis element $1$, with respect to the basis $T$, is an algebra homomorphism. Thus $\hat\bos^0$ has invariant $\ep$-rank, by theorem 1.

The three other cases are similar.\hfill QED.
\smallbreak

\begin{theorem} 
Let $B^h$ be one of the algebras of Theorem 2 (or more generally, any formal deformation of an $\ep$-commutative algebra), and call $\ep$ the corresponding commutation factor. Let us define :
$$\forall_h\ x,y\in B^0,\quad \{x,y\}_\ep=\mu_1(x,y)-\ep(\bar x,\bar y)\mu_1(y,x)$$
Then, $(B^0,\{.,.\}_\ep)$ is an $\ep$-Poisson algebra.
\end{theorem}
\dem\newline
If we write $[.,.]_\ep$ for the $\ep$-commutator in $(B^0,\mu^h)$, $\ep$-commutativity of $\mu_0$ implies :
$$[x,y]_\ep=\hbar\{x,y\}_\ep+{\cal O}(\hbar^2)$$
where ${\cal O}(\hbar^2)$ stands for terms of order $\geq\hbar^2$. Thus, it is clear that $\{.,.\}_\ep$ defines an $\ep$-Poisson algebra structure on $B^0$.\hfill QED.

\underline{Remark} : Let us go back to the physical case. If we define :
$$p_i:={1\over\sqrt 2}(\ai+\aix)$$
$$q_i:={1\over i\sqrt 2}(\aix-\ai)$$
and 
$$H_i:=hN_i$$
we find that all the equations for a system of $\ep$-classical harmonic oscillators are satisfied : 
$$\{p_i,p_j\}_\ep=\{q_i,q_j\}_\ep=0$$
$$\{p_i,q_j\}_\ep=\delta_{ij}$$
$$\{H_i,p_j\}_\ep=\delta_{ij}q_j$$
$$\{H_i,q_j\}_\ep=-\delta_{ij}p_j$$


\end{document}